\documentclass{ws-mplb}

\usepackage[super]{cite}
\usepackage{graphicx}
\usepackage{amsmath,amssymb}
\usepackage{mathtools}
\usepackage[mathscr]{eucal}
\usepackage{dsfont}

\providecommand{\ket}[1]{\left\lvert #1\right\rangle}%
\providecommand{\bra}[1]{\left\langle #1\right\rvert}%
\providecommand{\braket}[2]{\left<#1\vphantom{#2}\right|\left.\!\!#2\vphantom{#1}\right>}%
\providecommand{\nket}[1]{\left\lvert #1\right\}}%
\providecommand{\nbra}[1]{\left\{ #1\right\rvert}%
\providecommand{\nbraket}[2]{\left\{ #1\vphantom{#2}\right|\left.\!\!#2\vphantom{#1}\right\}}%
\providecommand{\nbracket}[3]{\left\{#1\vphantom{#2#3}\right|#2\left|#3\vphantom{#1#2}\right\}}%

\providecommand{\lbra}[1]{\prescript{}{\mathrm{L}}{\!\left\langle #1\right\rvert}}%
\providecommand{\rket}[1]{\left\lvert #1\right\rangle_\mathrm{R}}%
\providecommand{\lrbraket}[2]{\prescript{}{\mathrm{L}}{\!\left\langle #1\vphantom{#2}\right\rvert\left.\!\!#2\vphantom{#1}\right\rangle_\mathrm{R}}}%
\providecommand{\lbraket}[2]{\prescript{}{\mathrm{L}}{\!\left\langle #1\vphantom{#2}\right\rvert\left.\!\!#2\vphantom{#1}\right\rangle}}%
\providecommand{\rbraket}[2]{\left\langle #1\vphantom{#2}\right\rvert\left.\!\!#2\vphantom{#1}\right\rangle_\mathrm{R}}%

\providecommand{\e}{\mathrm{e}}
\providecommand{\I}{\mathrm{i}}
\DeclareMathOperator{\tr}{tr}

\begin{document}

\markboth{R. B. B. Santos and V. R. da Silva}
{Non-Hermitian Model for Asymmetric Tunneling}

%
\catchline{}{}{}{}{}
%

\title{NON-HERMITIAN MODEL FOR ASYMMETRIC TUNNELING}

\author{\footnotesize ROBERTO B. B. SANTOS\footnote{Corresponding author.}}

\address{Physics Department, Centro Universit\'{a}rio da FEI,\\ Av. Humberto de A. C. Branco 3972,
09850-901, S\~{a}o Bernardo do Campo, SP, Brazil\\
rsantos@fei.edu.br}

\author{\footnotesize VIN\'{I}CIUS R. DA SILVA}

\address{Electrical Engineering Department, Centro Universit\'{a}rio da FEI,\\ Av. Humberto de A. C. Branco 3972, 09850-901, S\~{a}o Bernardo do Campo, SP, Brazil\\
rocha\_vinicius.s@terra.com.br}

\maketitle

\begin{history}
\received{(Day Month Year)}
\revised{(Day Month Year)}
\end{history}

\begin{abstract}
We present a simple non-hermitian model to describe the phenomenon of asymmetric tunneling between two energy-degenerate sites coupled by a non-reciprocal interaction without dissipation. The system was described using a biorthogonal family of energy eigenvectors, the dynamics of the system was determined by the Schr\"{o}dinger equation, and unitarity was effectively restored by proper normalization of the state vectors. The results show that the tunneling rates are indeed asymmetrical in this model, leading to an equilibrium that displays unequal occupation of the degenerate systems even in the absence of external interactions.

\end{abstract}


\section{Introduction}

Tunneling has been one of the most characteristic features of quantum mechanics since the pioneering works by Hund~\cite{Hund:ZPhys1927} on tunneling in atoms and molecules, by Nordheim~\cite{Nordheim:ZPhys1927} and Oppenheimer~\cite{Oppenheimer:PR1928} on tunneling into the continuous, and by Gamow~\cite{Gamow:ZPhys1928, Gamow:Nature1928}, by Gurney and Condon~\cite{Gurney+Condon:Nature1928}, and by Born~\cite{Born:ZPhys1929} with its application to $\alpha$ particle decay. It should be noticed that the outgoing wavefunction method of Gamow leads to eigenfunctions with complex eigenvalues, which may be expressed as $E_0 - \I\Gamma$, where $\Gamma>0$ is proportional to the decay constant of the state. Hence, Gamow method is equivalent to the introduction of a non-hermitian effective hamiltonian to describe the system. Although non-hermitian hamiltonians are commonly used in scattering theory~\cite{Okolowicz+Ploszajczak+Rotter:PhysRep2003}, standard textbook-like tunneling between sites $A$ and $B$ is always symmetric owing to the hermitian character of the hamiltonian and to the fact that the density of states for the initial and for the final states are the same. Several aspects of the tunneling theory are reviewed, for instance, in the book by Razavy~\cite{Razavy:2003}.

Various signs of asymmetric tunneling rates were observed in the last $15$ years, however, in systems as diversified as high-$T_\mathrm{c}$ superconductors~\cite{Wei+etal:PRB1998, Wu+Koshizuka+Tanaka:MPLB1999, Pan+etal:Nature2000, Pan+etal:Nature2001}, non-superconducting cuprates~\cite{Hanaguri+etal:Nature2004}, Bose-Einstein condensates trapped in optical potentials~\cite{Wu+Niu:PRA2000, Jona-Lasinio+etal:PRL2003, Jona-Lasinio+etal:LaserPhys2005}, molecular electronics~\cite{Kornilovitch+etal:PRB2002, Elbing+etal:PNAS2005}, and, more recently, quantum dots interacting with leads~\cite{Rogge+etal:PhysicaE2006}, for instance. Despite the diversity of the physical mechanisms responsible for asymmetric tunneling in any of these systems, the unifying trace is that tunneling between initial and final sites involves a coupling to an external system, characterizing an open system. Hence, it is fair to say that asymmetric tunneling must be a feature present in several real systems.

Since the hermiticity of the hamiltonian plays a part in guaranteeing the symmetry of the tunneling, could a non-hermitian hamiltonian be capable of describing asymmetric tunneling in a simple way?

In a recent letter~\cite{Santos:EPL2012}, we presented a non-hermitian quantum model to describe the tunneling, through a $2D$-chiral mirror, of the light fields of a system of two resonant cavities in an attempt to understand the $4\,\mathrm{dB}$ asymmetry observed earlier in the transmission of circularly polarized photons through an array of asymmetric split rings~\cite{Plum+Fedotov+Zheludev:2011}. The model consisted simply in a pair of quantum oscillators coupled by a non-reciprocal interaction, and it was recently generalized to include number non-conservation and nonlinear interactions~\cite{Karakaya+etal:EPL2014}. Based on this simple model, hermitian and non-hermitian dynamics of mode entanglement between the light fields in the cavities were analyzed~\cite{Hardal:arXiv:1405.5079v3}.

We remark that the model incorporated phenomenologically the non-trivial interactions between photons and the chiral mirror without a detailed treatment of the mirror. In fact, there was no degrees of freedom corresponding to the mirror in the hamiltonian, and the effect of the mirror on the dynamics was introduced phenomenologically in the model in the non-reciprocal character of the interaction between the cavities. These works hint at the suggestion that non-hermitian quantum models are useful to describe certain classes of open systems in a simplified way.

Interest in non-hermitian quantum mechanics was renewed after Bender and Boettcher~\cite{Bender+Boettcher:1998} presented a class of non-hermitian hamiltonians which were symmetrical under spacetime inversion ($PT$-symmetry), and which exhibited real spectra. Could  $PT$ symmetry alone be responsible~\cite{Bender+Boettcher+Meisinger:JMP1999} for both the reality of the spectra and the unitarity of the theory? 

Mostafazadeh~\cite{Mostafazadeh:JMP2002, Mostafazadeh:2003b} showed that $PT$ symmetry was neither a necessary nor a sufficient condition for a hamiltonian to exhibit a real spectrum, and that exact $PT$ symmetry was equivalent to hermiticity with respect to a suitably defined inner product. Recent reviews~\cite{Bender:2007, Geyer+Heiss+Scholtz:CanJPhys2008, Mostafazadeh:2009} and the references therein provide a detailed exposition of the current status of the non-hermitian version of quantum mechanics.

Coupled systems were introduced in an analysis of optical $PT$-symmetric structures with a balanced gain-loss profile~\cite{Ruschhaupt+Delgado+Muga:JPhysA2004, El-Ganainy+etal:OptLett2007, Joglekar:EPJ2013}, and the first experimental realizations of such systems appeared shortly after that~\cite{Klaiman+Moiseyev+Gunther:2008, Guo+etal:2009, Ruter+etal:2010, Schindler+etal:2011}. In all these systems, non-hermiticity was caused by the reciprocal, balanced interaction between a lossy and an active element. It should be noticed, however, that  these pioneer experiments were classical emulations of the quantum $PT$-symmetric theory owing to the equivalence between the electromagnetic wave equation and the Schr\"{o}dinger equation for certain regimes.

In this paper, we analyze kinematical and dynamical aspects of a non-hermitian model describing tunneling between two sites coupled by a non-reciprocal interaction. The kinematical part consists in discussing the representation of the system in a biorthogonal family of energy eigenvectors, owing to the non-hermiticity of the hamiltonian, and correcting for the lack of strict unitarity of the theory. The dynamical part consists in computing the time evolution of the state vector describing the system in order to determine transition probabilities.

We emphasize that a novel aspect of this model is that the source of non-hermiticity is the presence of a non-reciprocal coupling between two otherwise ordinary quantum systems. An advantage of this treatment is that it allows the description of some aspects of the behavior of an open system using a formalism reminiscent of that applied to closed systems, that is, wavefunctions. While a closed system may be treated by the wavefunction or by the density operator formalisms, open systems are described by a density operator obeying a master equation. However, since the number of elements of the density matrix is the square of the number of coefficients of a wavefunction, solving the master equation for the density operator may be a difficult task.

Therefore, we propose to describe the dynamics of an open system in a way that may be simpler than the usual master equation method at the cost of using a non-hermitian hamiltonian. Hence, the approach is a phenomenological one, in which we begin with a non-hermitian hamiltonian and work out its consequences instead of attaining it from first principles in a more rigorous, yet more complex, open systems formalisms such as the Feshbach projection method~\cite{Feshbach:AnnPhys1958, Rotter:JPhysA2009}, the adiabatic elimination~\cite{Brion+Pedersen+Molmer:JPhysA2007, Reiter+Sorensen:PRA2012}.

\section{Brief description of the model}
\label{sec:Description}

\begin{figure}[htb]
 \centering
 \includegraphics[clip]{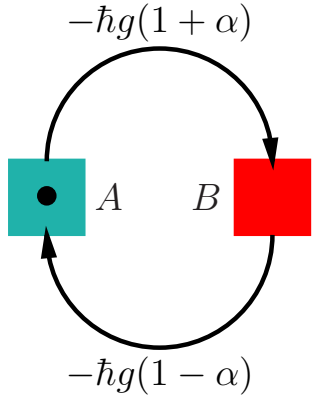}
 \caption{Illustration of the model. The excitation is localized at site $A$.}
 \label{fig:figure1}
\end{figure}

In this section, we describe a non-hermitian model for the tunneling of an excitation between two sites, $A$ and $B$, and coupled by the non-reciprocal interaction
\begin{equation}\label{eq:V}
  H= -\hbar g\left((1+\alpha)\sigma_\mathrm{A}^-\sigma_\mathrm{B}^+ + (1-\alpha)\sigma_\mathrm{A}^+\sigma_\mathrm{B}^- \right)
\end{equation}
where $\sigma_i^s$ is a standard two-state ladder (raising, $s=+$; lowering, $s=-$) operator for site $i$, $g$ is a real coupling constant, and $-1< \alpha <1$ is a parameter measuring the non-reciprocality of the interaction and the non-hermiticity of the hamiltonian. Since each site may be in an occupied or in an unoccupied state, it is natural to represent them by fermionic two-state operators. If $\alpha\neq 0$, the hamiltonian will be a non-hermitian operator owing to the non-reciprocal character of the interaction between the two sites as illustrated in Figure~\ref{fig:figure1}.

The interaction $H$ is represented as
\begin{equation}\label{eq:V-matrix-4states}
 H = -\hbar g
 \begin{bmatrix}
 0 & 0        & 0        & 0 \\
 0 & 0        & 1-\alpha & 0 \\
 0 & 1+\alpha & 0        & 0 \\
 0 & 0        & 0        & 0
 \end{bmatrix}
\end{equation}
relative to the basis formed by the vectors $\ket{11}$, $\ket{10}$, $\ket{01}$, and $\ket{00}$. The basis vectors are simultaneous eigenvectors of occupation number operators $N_\mathrm{A}\otimes\mathds{1}_\mathrm{B}$ and $\mathds{1}_\mathrm{A}\otimes N_\mathrm{B}$ for sites $A$ and $B$, respectively, where
\begin{equation}\label{eq:excitation_operator}
 N_i = \frac{\sigma_i^z + \mathds{1}_i}{2}.
\end{equation}

Since we are interested in single-excitation tunneling between the two sites, we will restrict ourselves to the single-excitation subspace spanned by the vectors $\ket{10}$ and $\ket{01}$ in order to determine the excitation exchange dynamics. In this case, the interaction is represented by
\begin{equation}\label{eq:V-matrix}
 H = -\hbar g
 \begin{bmatrix}
 0        & 1-\alpha \\
 1+\alpha & 0    \\
 \end{bmatrix}.
\end{equation}
Formally, this operator is a two-state non-hermitian hamiltonian given by $H = H_0 + \I V$ where $H_0=-\hbar g \sigma_x$ and the traceless $V = -\hbar g\alpha\sigma_y$ are both hermitian operators. This hamiltonian is similar, although not identical, to the one studied by Sergi and Zloshchastiev in the context of the open system dynamics of a single two-state system~\cite{Sergi+Zloshchastiev:IJMPB2013}. In their case, however, $\tr V <0$, which leads to dissipation, loss of quantum coherence and probability non-conservation.

Two-state hamiltonians having the generic form $H= H_0 + \I V$ where $H_0$ and $V$ are both hermitian operators were presented and discussed earlier in the context of the quantum brachistochrone problem~\cite{Bender+Brody+Jones+Meister:PRL2007, Mostafazadeh:PRL2007, Gunther+Samsonov:PRL2008}, or in the context of the closely related no-signaling condition violation polemic~\cite{Lee+Hsieh+Flammia+Lee:PRL2014, Znojil:arXiv:1404.1555v1}.

Besides, it must be pointed out that the no-signaling condition violation polemic refers exclusively to the status of $PT$-symmetric quantum theory as a fundamental description of nature. It is undisputable that non-hermitian models are useful even if they are only effective, phenomenological theories, as used successfully in several recent cases~\cite{Hiller+Kottos+Ossipov:PRA2006, Graefe+Korsch+Niederle:PRL2008, Rotter:JPhysA2009, Garcia-Calderon+Mattar+Villavicencio:PhysScr2012, Rapedius+Korsch:PRA2012, Wimberger+Parra-Murillo+Kordas:JPhysConfSer2013}.

\section{Kinematics and dynamics of the non-hermitian system}
\label{sec:kinematics+dynamics}

The non-reciprocal character of the interaction leads to a non-symmetrical, non-hermitian matrix representing the hamiltonian. A non-hermitian matrix has two sets of eigenvectors associated with its eigenvalues~\cite{Scholtz+Geyer+Hahne:AnnPhys1992, Moiseyev:2011}. The right eigenvectors are determined by
\begin{equation}\label{eq:right_eigenequation}
 H\rket{E} = E \rket{E}
\end{equation}
while the left eigenvectors are determined by
\begin{equation}
  \lbra{E} H = E \lbra{E}
\end{equation}
For this simple $2\times 2$ matrix, it is very easy to see that the two eigenvalues of $H$ are $E^\pm = \pm\hbar\omega$, with $\omega=|g|\sqrt{1-\alpha^2}$, and that the right eigenvectors are
\begin{equation}
 \rket{E^\pm} = \frac{1}{\sqrt{2}}
 \begin{bmatrix}
  1/\sqrt{\beta} \\
  \mp 1
 \end{bmatrix}
\end{equation}
while the left eigenvectors are
\begin{equation}
 \lbra{E^\pm} = \frac{1}{\sqrt{2}}
 \begin{bmatrix}
  \sqrt{\beta} & \mp 1
 \end{bmatrix},
\end{equation}
with $\beta=(1+\alpha)/(1-\alpha)$. These eigenvectors were normalized so that the completeness relation reads
\begin{equation}\label{eq:completeness}
 \sum_{s=\pm} \rket{E^s}\lbra{E^s} = \mathds{1},
\end{equation}
and the pair of the sets of right and of left eigenvectors forms a biorthogonal system, with
\begin{equation}
 \lrbraket{E^r}{E^s} = \delta^{rs}.
\end{equation}

For completeness, we would like to present some aspects regarding the usage of biorthogonal bases to represent quantum systems, since they seldom appear in quantum mechanics. We will begin with the representation of a state with respect to the energy biorthogonal bases. Using the completeness relation, Equation~\eqref{eq:completeness}, a state $\ket{\psi}$ may be represented by
\begin{equation}\label{eq:ket_representation}
 \ket{\psi} = \sum_s \rket{E^s} \lbraket{E^s}{\psi} = \sum_s c_\mathrm{L}^s \rket{E^s}
\end{equation}
where the expansion coefficients are given by $c_\mathrm{L}^s = \lbraket{E^s}{\psi}$. Similarly, $\bra{\psi}$ is represented by
\begin{equation}\label{eq:bra_representation}
 \bra{\psi} = \sum_s \rbraket{\psi}{E^s}\lbra{E^s} = \sum_s c_\mathrm{R}^s \lbra{E^s}
\end{equation}
with $c_\mathrm{R}^s = \rbraket{\psi}{E^s}$. Hence, there are two sets of expansion coefficients, one for kets, another for bras. As we will see next, these coefficients are not simply the complex conjugates of each other, as they would be if the hamiltonian were a hermitian operator.

Regarding time evolution, we assume that Schr\"{o}dinger equation
\begin{equation}\label{eq:Schrodinger_ket}
 \I\hbar\partial_t \ket{\psi(t)} = H\ket{\psi(t)}
\end{equation}
is still valid. Hence, using Eqs.~\eqref{eq:right_eigenequation}, \eqref{eq:ket_representation}, and~\eqref{eq:Schrodinger_ket}, we find that
\begin{equation}\label{eq:c_L(t)}
 c_\mathrm{L}^s (t) = c_\mathrm{L}^s (0) \e^{-\I E^s t/\hbar}.
\end{equation}
Although the coefficients $c_\mathrm{L}^s$ are similar to the familiar expansion coefficients obtained with a hermitian hamiltonian, the evolution is still non-unitary owing to the non-hermiticity of the hamiltonian operator. In order to grasp this point, we turn our attention to the Schr\"{o}dinger equation for $\bra{\psi(t)}$,
\begin{equation}\label{eq:Schrodinger_bra}
 -\I\hbar\partial_t \bra{\psi(t)} = \bra{\psi(t)} H^\dagger.
\end{equation}
Since $\lbra{E^s} H^\dagger \neq E^s \lbra{E^s}$, the expansion coefficients $c_\mathrm{R}^s$ for $\bra{\psi(t)}$ will not have a form as simple as that of the coefficients $c_\mathrm{L}^s$ for $\ket{\psi(t)}$. In fact, we find that
\begin{align}
 \dot{c}_\mathrm{R}^+ + \dot{c}_\mathrm{R}^- &= \I g(1+\alpha)\beta^{-3/2} \left(c_\mathrm{R}^+ - c_\mathrm{R}^- \right)  \\
 \dot{c}_\mathrm{R}^+ - \dot{c}_\mathrm{R}^- &= \I g(1-\alpha)\beta^{3/2} \left(c_\mathrm{R}^+ + c_\mathrm{R}^- \right)
\end{align}
which may be solved, after a slightly tedious algebra, to give
\begin{equation}\label{eq:c_R(t)}
  c_\mathrm{R}^\pm (t) =  \pm\frac{\I}{2\beta}\Bigl(\bigl(1+\beta^2\bigr)c_\mathrm{R}^\pm (0) -\bigl(1-\beta^2\bigr)c_\mathrm{R}^\mp (0) \Bigr)\sin(\omega t) + c_\mathrm{R}^\pm (0) \cos(\omega t).
\end{equation}

At this point, it should be clear that neither $\sum_s |c_\mathrm{L}^s|^2$ nor $\sum_s |c_\mathrm{R}^s|^2$ can represent the normalization of a state. In fact, even the quantity
\begin{equation}\label{eq:tentative_normalization}
 \braket{\psi}{\psi} = \sum_s c_\mathrm{L}^s c_\mathrm{R}^s
\end{equation}
is not the right choice, since it does not preserve the norm of the state as the system evolves. Suppose that the initial state is properly normalized, i.e., that $\sum_s c_\mathrm{L}^s (0) c_\mathrm{R}^s (0) = 1$. Then, Equations~\eqref{eq:c_L(t)}, and~\eqref{eq:c_R(t)} show that, in this case, the normalization factor is given by
\begin{multline}
 \braket{\psi}{\psi} = c_\mathrm{L}^+ (0) \e^{-\I \omega t}\biggl\{c_\mathrm{R}^+ (0)\cos(\omega t) + \frac{\I}{2\beta}\biggl[ c_\mathrm{R}^+ (0)\Bigl(1+\beta^2 \Bigr) - c_\mathrm{R}^- (0)\Bigl(1-\beta^2\Bigr) \biggr]\sin(\omega t) \biggr\} + \\
    c_\mathrm{L}^- (0) \e^{\I \omega t}\biggl\{c_\mathrm{R}^- (0)\cos(\omega t) + \frac{\I}{2\beta}\biggl[ c_\mathrm{R}^+ (0)\Bigl(1-\beta^2 \Bigr) - c_\mathrm{R}^- (0)\Bigl(1+\beta^2\Bigr) \biggr]\sin(\omega t) \biggr\}
\end{multline}
which is not a constant, in general. However, we can fix this unitarity problem by assuming that the proper normalized vectors are given by
\begin{align}\label{eq:normalization-i}
  \nket{\psi(t)} &= \frac{\ket{\psi(t)}}{\sqrt{\braket{\psi(t)}{\psi(t)}}} \\
  \nbra{\psi(t)} &= \frac{\bra{\psi(t)}}{\sqrt{\braket{\psi(t)}{\psi(t)}}},
\label{eq:normalization-f}
\end{align}
that is, the unnormalized vector multiplied by the inverse of the normalization factor, which is consistent with the prescription presented earlier by Scholtz, Geyer, and Hahne~\cite{Scholtz+Geyer+Hahne:AnnPhys1992}. An alternative method to define a unitary quantum system using a non-hermitian hamiltonian was presented by Mostafazadeh~\cite{Mostafazadeh:IJGMMP2010}, consisting in redefining the inner product of the theory with a metric operator, which, in our case, would be given by
\begin{equation}
 \eta_+ = \sum_s \ket{E^s}_\mathrm{L} \lbra{E^s} =
 \begin{bmatrix}
 \beta  &  0  \\
    0   &  1
 \end{bmatrix}.
\end{equation}
We believe that the simple method outlined above is sufficient for our purposes.

\section{Transition amplitudes and probabilities}
\label{sec:amplitudes+probabilities}

Having discussed these points, we finally turn our attention to probability amplitudes, transition probabilities, and normalization. Since the vector $\ket{\psi(t)}$ is not properly normalized as time passes, we define an unnormalized probability amplitude for the system initially in the state $\ket{\psi(0)}$ to be found in the state $\ket{\varphi}$ at time $t$ as the usual
\begin{equation}
 \mathscr{A}_{\psi\rightarrow\varphi}(t) = \braket{\varphi}{\psi(t)} = \sum_s d_\mathrm{R}^s c_\mathrm{L}^s (t)
\end{equation}
where $d_\mathrm{R}^s = \rbraket{\varphi}{E^s}$ are the expansion coefficients of $\bra{\varphi}$, and $\ket{\psi(t)}$ is the time evolved form of the initial state $\ket{\psi(0)}$. However, using our prescription for normalizing a state, Equations~\eqref{eq:normalization-i} and~\eqref{eq:normalization-f}, the normalized probability amplitude is given by
\begin{equation}
 \mathscr{A}_{\psi\rightarrow\varphi}^\mathrm{norm}(t) = \nbraket{\varphi}{\psi(t)} = \frac{\braket{\varphi}{\psi(t)}}{\sqrt{\braket{\varphi}{\varphi}}\sqrt{\braket{\psi(t)}{\psi(t)}}}.
\end{equation}
In this formalism, normalized transition probabilities would be given by
\begin{equation}
 \mathscr{P}_{\psi\rightarrow\varphi}^\mathrm{norm}(t) = \left|\mathscr{A}_{\psi\rightarrow\varphi}^\mathrm{norm}(t) \right|^2.
\end{equation}

In order to show that tunneling is indeed asymmetrical in this system, we focus on two different initial conditions: one in which the initial excitation is localized in site $A$, the other in which the initial excitation is in site $B$. In the former case, the initial state is given by
\begin{equation}
 \ket{A} = \ket{\psi_\mathrm{A}(0)} =
 \begin{bmatrix}
 1 \\
 0
 \end{bmatrix}
\end{equation}
while it is
\begin{equation}
 \ket{B} = \ket{\psi_\mathrm{B}(0)} =
 \begin{bmatrix}
 0 \\
 1
 \end{bmatrix}
\end{equation}
in the latter case. Expansion coefficients are easy to calculate:
\begin{align}
\label{eq:coefficients-i}
 c_\mathrm{LA}^- (0) &= \frac{\beta^{1/2}}{\sqrt{2}} = c_\mathrm{LA}^+ (0) \\
 c_\mathrm{RA}^- (0) &= \frac{\beta^{-1/2}}{\sqrt{2}} = c_\mathrm{RA}^+ (0) \\
 c_\mathrm{LB}^- (0) &= \frac{1}{\sqrt{2}} = -c_\mathrm{LB}^+ (0) \\
 c_\mathrm{RB}^- (0) &= \frac{1}{\sqrt{2}} = -c_\mathrm{RB}^+ (0),
\label{eq:coefficients-f}
\end{align}
and one may verify that the initial states are properly normalized. It is also very easy to verify that the two initial states are orthogonal since
\begin{equation}
 \braket{\psi_\mathrm{B}(0)}{\psi_\mathrm{A}(0)} = \sum_s c_\mathrm{RB}^s (0) c_\mathrm{LA}^s (0) = 0.
\end{equation}
With all these coefficients, it is possible to calculate the normalization factors. For $\ket{\psi_\mathrm{A}(t)}$, we have
\begin{equation}
  \braket{\psi_\mathrm{A}(t)}{\psi_\mathrm{A}(t)} = \cos^2(\omega t) + \beta\sin^2(\omega t)
\end{equation}
while the normalization factor for $\ket{\psi_\mathrm{B}(t)}$ is slightly different, being given by
\begin{equation}
  \braket{\psi_\mathrm{B}(t)}{\psi_\mathrm{B}(t)} = \cos^2(\omega t) + \beta^{-1}\sin^2(\omega t).
\end{equation}

In each case, there are two unnormalized probability amplitudes that are interesting to calculate, the amplitudes for an excitation to be or not to be exchanged between sites. When the excitation is initially at site $A$, the former amplitude is
\begin{equation}
  \mathscr{A}_{A\rightarrow B} = \braket{B}{\psi_\mathrm{A}(t)} = \sum_s c_\mathrm{RB}^s (0) c_\mathrm{LA}^s (t) = \I\beta^{1/2}\sin(\omega t),
\end{equation}
 and the latter amplitude is given by
\begin{equation}
  \mathscr{A}_{A\rightarrow A} = \braket{A}{\psi_\mathrm{A}(t)} = \sum_s c_\mathrm{RA}^s (0) c_\mathrm{LA}^s (t) = \cos(\omega t).
\end{equation}
However, when the excitation is initially at site $B$, the amplitudes are given by
\begin{align}
 \mathscr{A}_{B\rightarrow A} &= \braket{A}{\psi_\mathrm{B}(t)} = \I\beta^{-1/2}\sin(\omega t), \\
 \mathscr{A}_{B\rightarrow B} &= \braket{B}{\psi_\mathrm{B}(t)} = \cos(\omega t)
\end{align}

\begin{figure}[htb]
 \centering
 \includegraphics[clip]{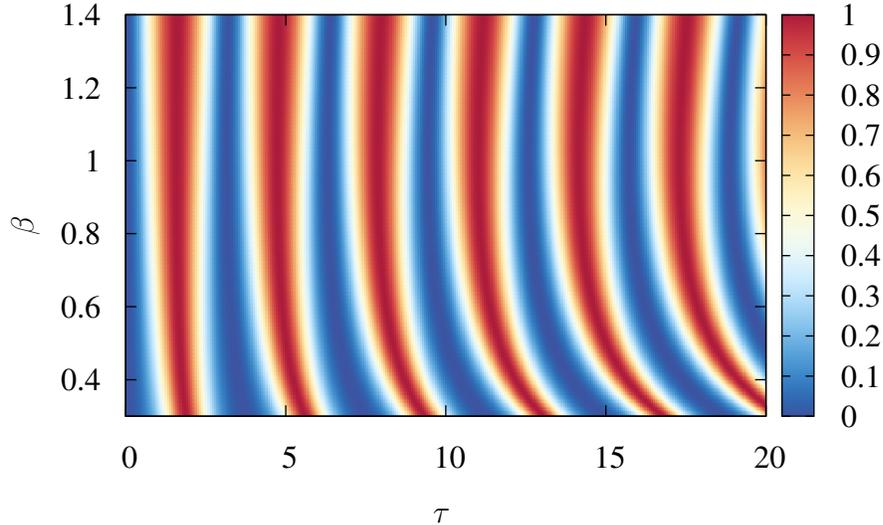}
 \caption{Image plot of the transition probability $\mathscr{P}_{A\rightarrow B}$ as a function of the parameters $\tau=|g| t$, and $\beta=(1+\alpha)/(1-\alpha)$.}
 \label{fig:figure2}
\end{figure}

Determination of all these amplitudes allows us to calculate the normalized transition probabilities. The probability for an excitation initially at site $A$ to be found at site $B$ at time $t$ is
\begin{equation}
 \mathscr{P}_{A\rightarrow B}^\mathrm{norm} (t) = \frac{\beta\sin^2(\omega t)}{\cos^2(\omega t) + \beta\sin^2(\omega t)}
\end{equation}
as shown in Figure~\ref{fig:figure2} as a function of the two adimensional parameters $\tau = |g| t$, and $\beta = (1+\alpha)/(1-\alpha)$. For ease of reference, the time intervals in which $\mathscr{P}_{A\rightarrow B}^\mathrm{norm} > 0.5$ will be called the high $A\rightarrow B$ transition probability phase, and the time intervals in which $\mathscr{P}_{A\rightarrow B}^\mathrm{norm} < 0.5$ will be the low $A\rightarrow B$ transition probability phase. In the reciprocal case ($\beta=1$), the system spends equal amounts of time in the low and in the high $A\rightarrow B$ transition probability phases. As $\beta$  departs from unity, the non-reciprocality of the interaction manifests itself. In the $\beta<1$ region, the system passes more than $50\%$ of the time in the low $A\rightarrow B$ transition probability phase, while for $\beta>1$, the opposite happens. In all cases, transition probability is a periodic function with period $\pi/2\omega$. For the inverse process, the transition probability is given by
\begin{equation}
 \mathscr{P}_{B\rightarrow A}^\mathrm{norm} (t) = \frac{\beta^{-1}\sin^2(\omega t)}{\cos^2(\omega t) + \beta^{-1}\sin^2(\omega t)},
\end{equation}
and an analysis similar to the above holds. From this analysis, we may hint that, in the long run, the occupation of the two sites should be distinct, as we will show briefly.

\begin{figure}[htb]
 \centering
 \includegraphics[clip]{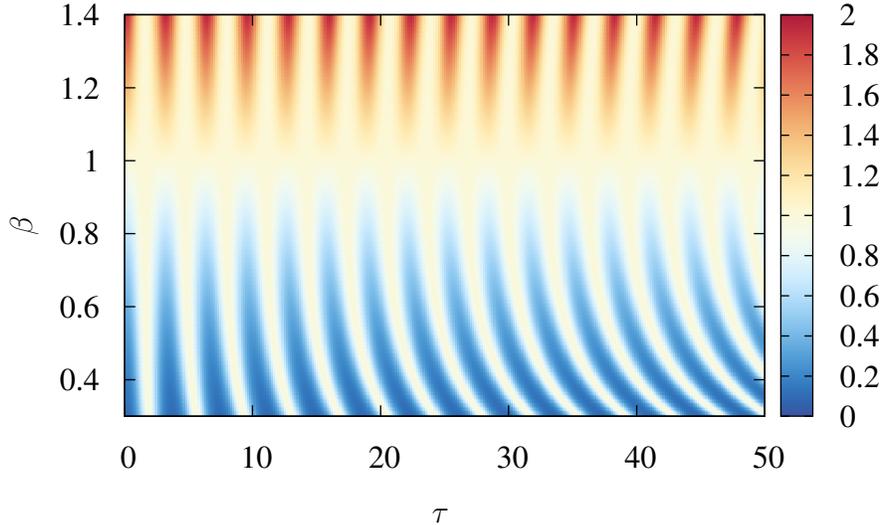}
 \caption{Image plot of the probability ratio $\mathscr{R}^{\underline{A\rightarrow B}}_{B\rightarrow A}$ as a function of the parameters $\tau=|g| t$, and $\beta=(1+\alpha)/(1-\alpha)$.}
 \label{fig:figure3}
\end{figure}

Figure~\ref{fig:figure3} shows the transition probability ratio
\begin{equation}
\mathscr{R}^{\underline{A\rightarrow B}}_{B\rightarrow A}=\frac{\mathscr{P}_{A\rightarrow B}^\mathrm{norm}}{ \mathscr{P}_{B\rightarrow A}^\mathrm{norm}} = \beta^2\frac{\cos^2(\omega t) + \beta^{-1}\sin^2(\omega t)}{\cos^2(\omega t) + \beta\sin^2(\omega t)}
\end{equation}
also as a function of the adimensional parameters $\tau$, and $\beta$. As expected, the maxima of the transition probability ratio depart from unity as $\beta$ becomes different from unity, signalling the asymmetry in the tunneling rates in this non-hermitian model.

Besides, the probability for an excitation initially at site $A$ still be found at site $A$ at time $t$ is
\begin{equation}
 \mathscr{P}_{A\rightarrow A}^\mathrm{norm} (t) = \frac{\cos^2(\omega t)}{\cos^2(\omega t) + \beta\sin^2(\omega t)},
\end{equation}
which means that
\begin{equation}
 \mathscr{P}_{A\rightarrow A}^\mathrm{norm} (t) + \mathscr{P}_{A\rightarrow B}^\mathrm{norm} (t) = 1.
\end{equation}
Similarly,
\begin{equation}
 \mathscr{P}_{B\rightarrow A}^\mathrm{norm} (t) + \mathscr{P}_{B\rightarrow B}^\mathrm{norm} (t) = 1,
\end{equation}
confirming that there is no loss in this model, after normalizing the states. However, dissipation may be added phenomenologically to the model using well-known techniques such as introducing a complex diagonal term in the hamiltonian matrix. Besides, it must be noticed that bypassing the normalization step leads, in general, to a non-unitary dynamics, that is,
\begin{equation}
\mathscr{P}_{i\rightarrow i} (t) + \mathscr{P}_{i\rightarrow i'} (t) \neq 1,
\end{equation}
which shows that the unnormalized description allows for excitation exchange with the environment.

In this regard, we would like to point out that the density operator for this system is given, in the unnormalized description, by
\begin{equation}
 \rho = \ket{\psi}\bra{\psi} = \sum_{r,s} c_\mathrm{L}^r c_\mathrm{R}^s \rket{E^r}\lbra{E^s}.
\end{equation}
Inspection of the Equations~\ref{eq:c_L(t)} and~\ref{eq:c_R(t)} for the wavefunction coefficients $c_\mathrm{L}^\pm$ and $c_\mathrm{R}^\pm$ is sufficient to show that there is no damping in the  quantum coherence in the dynamics of this system, even in the unnormalized case. This stands in opposition with the master equation approach to open quantum systems, in which normalization is preserved but coherence is rapidly lost as the system interacts with the environment~\cite{Caldeira+Leggett:PRA1985, Walls+Milburn:PRA1985}. In spite of this, long-lived coherence is observed even in very noisy environments, during photosynthesis~\cite{Engel+etal:Nature2007}, for instance. Since preservation of quantum coherence would be an important achievement in quantum information theory, it represents an advantage to know that an open system may slow down decoherence if its effective non-hermitian hamiltonian is similar to the one described here.

\begin{figure}[h!]
 \centering
 \includegraphics[clip]{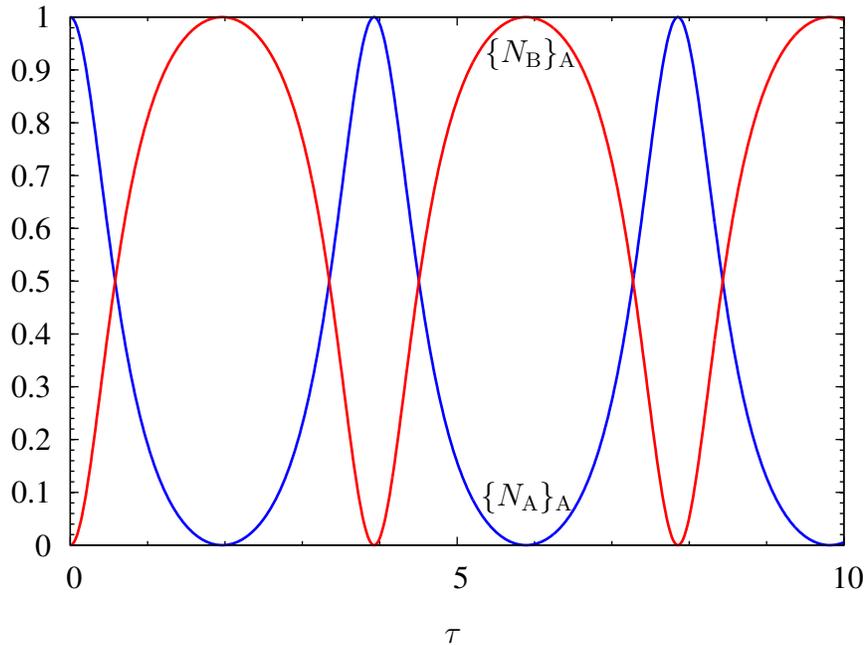}
 \caption{Occupation numbers $\{N_\mathrm{A}\}_\mathrm{A}$, and $\{N_\mathrm{B}\}_\mathrm{A}$ as functions of the adimensional parameter $\tau=|g| t$ for $\beta=4.0$ ($\alpha=0.6$). Excitation was initially at site $A$.}
 \label{fig:figure4}
\end{figure}

Directing our attention to the occupation number $\{N_\mathrm{A}\}_\mathrm{A} = \nbracket{\psi_\mathrm{A} (t)}{N_\mathrm{A}}{\psi_\mathrm{A} (t)}$ for site $A$, we find that
\begin{equation}
   \{N_\mathrm{A}\}_\mathrm{A} = \frac{\cos^2(\omega t)}{\cos^2(\omega t) + \beta\sin^2(\omega t)}
\end{equation}
when the excitation is initially at site $A$, for instance. In this same case, the occupation number $\{N_\mathrm{B}\}_\mathrm{A} = \nbracket{\psi_\mathrm{A} (t)}{N_\mathrm{B}}{\psi_\mathrm{A} (t)}$ for site $B$ is given by
\begin{equation}
 \{N_\mathrm{B}\}_\mathrm{A} = \frac{\beta\sin^2(\omega t)}{\cos^2(\omega t) + \beta\sin^2(\omega t)}.
\end{equation}
Figure~\ref{fig:figure4} shows the occupation numbers for the two sites plotted against the parameter $\tau=|g| t$ for a typical $\beta > 1$ case when the excitation was initially at site $A$. A distinct feature of the graph is that the occupation number for site $A$ is, on average, lower than the occupation number for site $B$. Clearly, in a $\beta < 1$ scenario, the opposite would happen.

\begin{figure}[h!]
\centering
\includegraphics[clip]{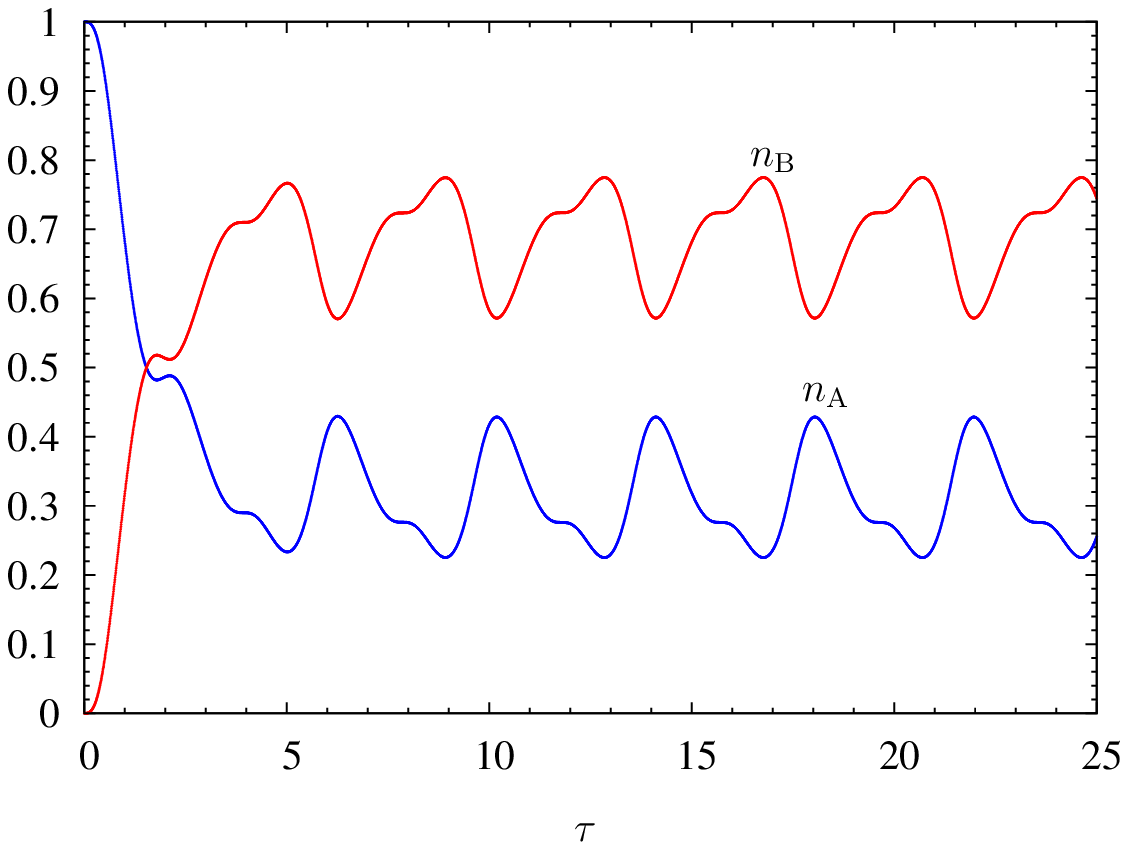}
\caption{Semiclassical occupation numbers $n_\mathrm{A}$, and $n_\mathrm{B}$ as functions of the adimensional parameter $\tau=|g| t$ for $\beta=4.0$ ($\alpha=0.6$). Excitation was initially at site $A$.}
\label{fig:figure5}
\end{figure}

Finally, having the transition probabilities, a semiclassical rate equation for this system is given simply by
\begin{equation}
 \dot{n}_\mathrm{A} = -n_\mathrm{A}\mathscr{P}_{A\rightarrow B}^\mathrm{norm} + n_\mathrm{B}\mathscr{P}_{B\rightarrow A}^\mathrm{norm} = -\dot{n}_\mathrm{B},
\end{equation}
since the excitation is not lost from the system. Figure~\ref{fig:figure5} shows the semiclassical occupation numbers $n_\mathrm{A}$ for site $A$, and $n_\mathrm{B}$ for site $B$ plotted against the parameter $\tau=|g| t$ for a typical $\beta > 1$ case when the excitation was initially at site $A$. After a short transient period, the semiclassical occupation numbers are clearly separated, fluctuating around their new equilibrium values. As expected in this $\beta > 1$ case, the occupation of site $B$ is greater than the occupation of site $A$. The opposite would happen in a $\beta <1$ case.

In conclusion, we have shown that a non-hermitian quantum model describes the excitation exchange dynamics typical of asymmetric tunneling. An effectively unitary description of this system was attained by proper normalization of the state vectors. A feature common to all non-hermitian systems is that the normalization depends on the hamiltonian, having to be chosen dynamically. However, even in the unnormalized case, quantum coherence is not exponentially destroyed. This treatment suggests a way of dealing phenomenologically with irreversible interacting quantum systems using a formalism that mimics that of a closed system, and which may be, therefore, simpler than the usual open system formalism.

\section*{Acknowledgments}

We acknowledge Dr.\! F. Altintas for fruitful discussions, two anonymous referees for criticisms, comments, and questions that really improved the paper, and Fr. Saboia de Medeiros Ignatian Educational Foundation for continuous support.


\begin{thebibliography}{0}

\bibitem{Hund:ZPhys1927} F. Hund, {\it Z. Phys.} {\bf 40}, 742 (1927).

\bibitem{Nordheim:ZPhys1927} L. Nordheim, {\it Z. Phys.} {\bf 46}, 833 (1927).

\bibitem{Oppenheimer:PR1928} J. R. Oppenheimer, {\it Phys. Rev.} {\bf 31}, 66 (1928).

\bibitem{Gamow:ZPhys1928} G. Gamow, {\it Z. Phys.} {\bf 51}, 204 (1928).

\bibitem{Gamow:Nature1928} G. Gamow, {\it Nature}, {\bf 122}, 805 (1928).

\bibitem{Gurney+Condon:Nature1928} R. W. Gurney and E. U. Condon, {\it Nature} {\bf 122}, 439 (1928).

\bibitem{Born:ZPhys1929} M. Born, {\it Z. Phys.} {\bf 58}, 306 (1929).

\bibitem{Okolowicz+Ploszajczak+Rotter:PhysRep2003} J. Okolowicz, M. Ploszajczak and I. Rotter, {\it Phys. Rep.} {\bf 374}, 271 (2003).

\bibitem{Razavy:2003} M. Razavy, {\it Quantum Theory of Tunneling} (World Scientific, Singapore, 2003).

\bibitem{Wei+etal:PRB1998} J. Y. T. Wei, C. C. Tsuei, P. J. M. van Bentum, Q. Xiong, C. W. Chu and M. K. Wu, {\it Phys. Rev. B} {\bf 57}, 3650 (1998).

\bibitem{Wu+Koshizuka+Tanaka:MPLB1999} A. T. Wu, N. Koshizuka and S. Tanaka, {\it Mod. Phys. Lett. B} {\bf 13}, 735 (1999).

\bibitem{Pan+etal:Nature2000} S. H. Pan, E. W. Hudson, K. M. Lang, H. Eisaki, S. Uchida and J. C. Davis, {\it Nature} {\bf 403}, 746 (2000).

\bibitem{Pan+etal:Nature2001} S. H. Pan, J. P. O'Neal, R. L. Badzey, C. Chamon, H. Ding,
J. R. Engelbrecht, Z. Wang, H. Eisaki, S. Uchida, A. K. Guptak,
K.-W. Ng, E. W. Hudson, K. M. Lang and J. C. Davis, {\it Nature} {\bf 413}, 282 (2001).

\bibitem{Hanaguri+etal:Nature2004} T. Hanaguri, C. Lupien, Y. Kohsaka, D.-H. Lee, M. Azuma, M. Takano, H. Takagi and J. C. Davis, {\it Nature} {\bf 430}, 1001 (2004).

\bibitem{Wu+Niu:PRA2000} B. Wu and Q. Niu, {\it Phys. Rev. A} {\bf 61}, 023402 (2000).

\bibitem{Jona-Lasinio+etal:PRL2003} M. Jona-Lasinio, O. Morsch, M. Cristiani, N. Malossi, J. H. M\"{u}ller, E. Courtade, M. Anderlini and E. Arimondo, {\it Phys. Rev. Lett.} {\bf 91}, 230406 (2003).

\bibitem{Jona-Lasinio+etal:LaserPhys2005} M. Jona-Lasinio, O. Morsch, M. Cristiani, E. Arimondo and C. Menotti, {\it Laser Phys.} {\bf 15}, 1180 (2005).

\bibitem{Kornilovitch+etal:PRB2002} P. E. Kornilovitch, A. M. Bratkovsky and R. S. Williams, {\it Phys. Rev. B} {\bf 66}, 165436 (2002).

\bibitem{Elbing+etal:PNAS2005} M. Elbing, R. Ochs, M. Koentopp, M. Fischer, C. von H\"{a}nisch, F. Weigend, F. Evers, H. B. Weber and M. Mayor, {\it Proc. Natl. Acad. Sci. USA} {\bf 102}, 8815 (2005).

\bibitem{Rogge+etal:PhysicaE2006} M. C. Rogge, C. Fricke, B. Harke, F. Hohls, R. J. Haug and W. Wegscheider, {\it Physica E} {\bf 34}, 500 (2006).

\bibitem{Santos:EPL2012} R. B. B. Santos, {\it EPL} {\bf 100}, 24005 (2012).

\bibitem{Plum+Fedotov+Zheludev:2011} E. Plum, V. A. Fedotov and N. I. Zheludev, {\it J. Opt.} {\bf 13}, 024006 (2011).

\bibitem{Karakaya+etal:EPL2014} E. Karakaya E., F. Altintas, K. G\"{u}ven and \"{O}. E. M\"{u}stecaplioglu, {\it EPL} {\bf 105}, 40001 (2014).

\bibitem{Hardal:arXiv:1405.5079v3} A. \"{U}. C. Hardal, arXiv:1405.5079v3 (2014).

\bibitem{Bender+Boettcher:1998} C. M. Bender and S. Boettcher, {\it Phys. Rev. Lett.} {\bf 80}, 5243 (1998).

\bibitem{Bender+Boettcher+Meisinger:JMP1999} C. M. Bender, S. Boettcher and P. N. Meisinger, {\it J. Math. Phys.} {\bf 40}, 2201 (1999).

\bibitem{Mostafazadeh:JMP2002} A. Mostafazadeh, {\it J. Math. Phys.} {\bf 43}, 205 (2002).

\bibitem{Mostafazadeh:2003b} A. Mostafazadeh, {\it J. Phys. A: Math. Gen.} {\bf 36}, 7081 (2003).

\bibitem{Bender:2007} C. M. Bender, {\it Rep. Prog. Phys.} {\bf 70}, 947 (2007).

\bibitem{Geyer+Heiss+Scholtz:CanJPhys2008} H. B. Geyer, W. D. Heiss and F. G. Scholtz, {\it Can. J. Phys.} {\bf 86}, 1195 (2008).

\bibitem{Mostafazadeh:2009} A. Mostafazadeh, {\it Pramana} {\bf 73}, 269 (2009).

\bibitem{Ruschhaupt+Delgado+Muga:JPhysA2004} A. Ruschhaupt, F. Delgado and J. G. Muga, {\it J. Phys. A: Math. Gen.} {\bf 38}, L171 (2005).

\bibitem{El-Ganainy+etal:OptLett2007} R. El-Ganainy, K. G. Makris, D. N. Christodoulides and Z. H. Musslimani, {\it Opt. Lett.} {\bf 32}, 2632 (2007).

\bibitem{Joglekar:EPJ2013} Y. N. Joglekar, C. Thompson, D. D. Scott and Gautam Vemuri, {\it Eur. Phys. J.: Appl. Phys.} {\bf 63}, 30001 (2013).

\bibitem{Klaiman+Moiseyev+Gunther:2008} S. Klaiman, N. Moiseyev and U. G\"{u}nther, {\it Phys. Rev. Lett.} {\bf 101}, 080402 (2008).

\bibitem{Guo+etal:2009} A. Guo, G. J. Salamo, D. Duchesne, R. Morandotti, M. Volatier-Ravat, V. Aimez, G. A. Siviloglou and D. N. Christodoulides, {\it Phys. Rev. Lett.} {\bf 103}, 093902 (2009).

\bibitem{Ruter+etal:2010} C. E. R\"{u}ter, K. G. Makris, R. El-Ganainy, D. N. Christodoulides, M. Segev and D. Kip, {\it Nature Physics} {\bf 6}, 192 (2010).

\bibitem{Schindler+etal:2011} J. Schindler, A. Li, M. C. Zheng, F. M. Ellis and T. Kottos, {\it Phys. Rev. A} {\bf 84}, 040101 (2011).

\bibitem{Feshbach:AnnPhys1958} H. Feshbach, {\it Ann. Phys.} {\bf 5}, 357 (1958).

\bibitem{Rotter:JPhysA2009} I. Rotter, {\it J. Phys. A: Math. Theor.} {\bf 42}, 153001 (2009).

\bibitem{Brion+Pedersen+Molmer:JPhysA2007} E. Brion, L. H. Pedersen and K. M{\o}lmer, {\it J. Phys. A: Math. Theor.} {\bf 40}, 1033 (2007).
    
\bibitem{Reiter+Sorensen:PRA2012} F. Reiter and A. S. S{\o}rensen, {\it Phys. Rev. A} {\bf 85}, 032111 (2012).

\bibitem{Sergi+Zloshchastiev:IJMPB2013} A. Sergi and K. G. Zloshchastiev, {\it Int. J. Mod. Phys. B} {\bf 27}, 1350163 (2013).

\bibitem{Bender+Brody+Jones+Meister:PRL2007} C. M. Bender, D. C. Brody, H. F. Jones and B. K. Meister, {\it Phys. Rev. Lett.} {\bf 98}, 040403 (2007).

\bibitem{Mostafazadeh:PRL2007} A. Mostafazadeh, {\it Phys. Rev. Lett.} {\bf 99}, 130502 (2007).

\bibitem{Gunther+Samsonov:PRL2008} U. G\"{u}nther and B. Samsonov, {\it Phys. Rev. Lett.} {\bf 101}, 230404 (2008).

\bibitem{Lee+Hsieh+Flammia+Lee:PRL2014} Y.-C. Lee, M.-H. Hsieh, S. T. Flammia and R.-K. Lee, {\it Phys. Rev. Lett.} {\bf 112}, 130404 (2014).

\bibitem{Znojil:arXiv:1404.1555v1} M. Znojil, arXiv:1404.1555v1 (2014).

\bibitem{Hiller+Kottos+Ossipov:PRA2006} M. Hiller, T. Kottos and A. Ossipov, {\it Phys. Rev. A} {\bf 73}, 063625 (2006).

\bibitem{Graefe+Korsch+Niederle:PRL2008} E. M. Graefe, H. J. Korsch and A. E. Niederle, {\it Phys. Rev. Lett.} {\bf 101}, 150408 (2008).

\bibitem{Garcia-Calderon+Mattar+Villavicencio:PhysScr2012} G. Garc\'{\i}a-Calder\'{o}n, A. M\'{a}ttar and J. Villavicencio, {\it Phys. Scr.} {\bf T151}, 01476 (2012).

\bibitem{Rapedius+Korsch:PRA2012} K. Rapedius and H. J. Korsch, {\it Phys. Rev. A} {\bf 86}, 025601 (2012).

\bibitem{Wimberger+Parra-Murillo+Kordas:JPhysConfSer2013} S. Wimberger, C. A. Parra-Murillo and G. Kordas, {\it J. Phys. Conf. Ser.}  {\bf 442}, 012029 (2013).

\bibitem{Scholtz+Geyer+Hahne:AnnPhys1992} F. G. Scholtz, H. B. Geyer and F. J. W. Hahne,
{\it Ann. Phys.} {\bf 213}, 74 (1992).

\bibitem{Moiseyev:2011} N. Moiseyev, {\it Non-Hermitian Quantum Mechanics} (Cambridge University Press, Cambridge, 2011).

\bibitem{Mostafazadeh:IJGMMP2010} A. Mostafazadeh, {\it Int. J. Geom. Meth. Mod. Phys.} {\bf 7}, 1191 (2010).

\bibitem{Caldeira+Leggett:PRA1985} A. O. Caldeira and A. J. Leggett, {\it Phys. Rev. A} {\bf 31}, 1059 (1985).
    
\bibitem{Walls+Milburn:PRA1985} D. F. Walls and G. J. Milburn, {\it Phys. Rev. A} {\bf 31}, 2403 (1985).

\bibitem{Engel+etal:Nature2007} G. S. Engel, T. R. Calhoun, E. L. Read, T.-K. Ahn, T. Mancal, Y.-C. Cheng, R. E. Blankenship and G. R. Fleming, {\it Nature} {\bf 446}, 782 (2007).

\end{thebibliography}
\end{document}